\documentclass[showpacs,preprintnumbers,amsmath,amssymb,aps,twocolumn]{revtex4}
\usepackage{graphicx}
\usepackage{epsf}
\usepackage{epsfig}

\newcommand{\step}{\vspace{.5em}}
\newcommand{\smallstep}{\vspace{.08em}}
\def\di{\displaystyle}

\def\bg{\begin{eqnarray}\begin{array}{rcl}\displaystyle}
\def\eg{\end{array} &\di    &\di   \end{eqnarray}}
\def\bm#1{\begin{eqnarray}\begin{array}{#1}\di}
\def\bmo#1{\begin{eqnarray*}\begin{array}{#1}\di}
\def\bml#1#2{\begin{eqnarray}\begin{array}{#1}\label{#2}\di}
\def\bgo{\begin{eqnarray*}\begin{array}{rcl}\displaystyle}
\def\ego{\end{array} &\di    &\di \nonumber  \end{eqnarray*}}

\def\btensor#1#2{\renew\left#1\begin{array}{#2}\di}
\def\brtensor#1#2#3{\ren#3\left#1\begin{array}{#2}}
\def\botensor#1#2{\renew\left#1\begin{array}{#2}}
\def\etensor#1{\end{array}\right#1}

\def\eq#1{(\ref{#1})}

\def\s0#1#2{\mbox{\small{$ \frac{#1}{#2} $}}}
\def\0#1#2{\frac{#1}{#2}}

\def\R{{{\rm l}\!{\rm R}}}

\def\ren#1{\renewcommand{\arraystretch}{#1}}

\def\renew{\renewcommand{\arraystretch}{1}}

\begin{document}

\title{Convexity of the effective action from functional flows}
\vspace{1.5 true cm}

\pacs{11.10.Gh,11.10.Hi,05.10.Cc\vspace{-.05cm}}
\author{Daniel F.~Litim ${}^{a,b}$}
\author{Jan M.~Pawlowski${}^{c}$}
\author{Lautaro Vergara${}^{d}$}
\address{ ${}^a$\mbox{School of Physics and Astronomy,
University of Southampton, Southampton SO17 1BJ, U.K.}\\
${}^b$Physics Department, CERN, Theory Division, CH-1211 Geneva 23.\\
${}^c$\mbox{ Institut
f{\"u}r Theoretische Physik, University of Heidelberg, 
Philosophenweg 16, 69120 Heidelberg,
Germany}\\
${}^d$\mbox{Departamento de
F\'{\i}sica, Universidad de Santiago de Chile, Casilla 307,
Santiago 2, Chile}}

 \preprint{HD-THEP-05-22, SHEP-0531, CERN-PH/TH-2005-252
}

\thispagestyle{empty}

\abstract{We show that convexity of the effective action follows from its
  functional flow equation. Our analysis is based on a new, spectral
  representation. The results are relevant for the study of physical
  instabilities. We also derive constraints for convexity-preserving
  regulators within general truncation schemes including proper-time flows,
  and bounds for infrared anomalous dimensions of propagators.}}

\maketitle

\pagestyle{plain}
\setcounter{page}{1}

\noindent {\it Introduction.---} Functional flows have been
successfully used for perturbative as well as non-perturbative
problems in quantum field theory and statistical physics
\cite{reviews,Litim:1998nf}. They provide a definition for finite
generating functionals of the quantum theory, $i.e.$ the effective
action. The latter is a Legendre transform and therefore convex
\cite{O'Raifeartaigh:1986hi}. In general, convex effective actions
admit stable solutions of the quantum equations of motions. In turn,
non-convexities are linked to instabilities and have physical as well
as technical origins. Physical instabilities range from those in
condensed matter systems to QCD instabilities and are $e.g.$~related
to tunnelling phenomena and decay properties \cite{reviews}. On the
other hand instabilities may reflect artefacts of the underlying
truncation or parameterisation. It is mandatory to properly
distinguish between these two qualitatively different
scenarios.\smallstep

Functional flows for the effective action have been constructed from
first principles as well as from a renormalisation group improvement.
A large class of the latter are well-defined truncations of
first-principle flows within a background field formulation
\cite{Litim:2001hk,Litim:2002xm}, including proper-time
flows \cite{SBLiao,SBLiaogauge,pirner}. For
first-principle flows, the set of convex functionals is an attractive
fixed point of the full flow.  Since truncations to the full problem
at hand are inevitable, it is vital to identify convexity-preserving
expansion schemes and regulators, and to determine limitations of
widely used approximation schemes. In this Letter we provide a
constructive proof of convexity for the effective action hence closing
the present conceptual gap. Throughout, we illustrate our reasoning at
the example of the derivative expansion. \step

\noindent {\it Functional flows and spectral representation.---} The
analysis is done within a new, spectral representation for functional
flows w.r.t.\ an infrared cutoff scale $k$,
\begin{equation}
  \partial_t\Gamma_k[\phi,\bar\phi]=\012 \int_\R 
  d\lambda\,
  \rho(\bar\phi;\lambda)
  \langle \psi_\lambda|\0{1}{
    \Gamma_k^{(2,0)}+R_k}\partial_t R_k|
  \psi_\lambda\rangle
\label{eq:specfullflow}\end{equation}
and $t=\ln k$. Here, $\phi$ is the dynamical quantum field and
$\bar\phi$ is some background configuration, $e.g.$~the vacuum field.
The flow \eq{eq:specfullflow} depends on the full propagator of the
quantum field $\phi$. The propagator is written in terms of the
two-point function $\Gamma^{(2,0)}$ of $\phi$. Generally we define
mixed functional derivatives w.r.t.\ $\phi$ and $\bar\phi$ as
$\Gamma_k^{(n,m)}=\delta^{n+m}\Gamma_k/(\delta\phi^n
\delta\bar\phi^m)$ \cite{Litim:2002xm}. The regulator
$R_k=R_k(\Gamma_k^{(2,0)}[\bar\phi,\bar\phi])$ depends on the
two-point function evaluated at the background field $\bar\phi$, and
the spectral values of $\Gamma_k^{(2,0)}$ are defined by
\begin{eqnarray}\label{eq:lambda}  
  \lambda(\phi,\bar\phi)=
  \langle\psi_\lambda |\Gamma_k^{(2,0)}[\phi,\bar\phi]|
  \psi_\lambda\rangle\,, 
\end{eqnarray}
with eigenfunctions $\psi_\lambda$, and $\rho(\bar\phi;\lambda)$ is
the spectral density of $\lambda(\bar\phi,\bar\phi)$.  The flow
\eq{eq:specfullflow} is fully equivalent to standard background field
flows studied in \cite{Litim:2002xm}. We note that, since
$\Gamma_k^{(2,0)}$ can have negative spectral values, $R_k>0$ also has
to be defined for negative arguments. In the absence of further scales
we write the regulator as $R_k(\lambda)=\lambda\, r(\lambda/k^2)$ with
$k$-independent function $r$.  As an example, consider
\eq{eq:specfullflow} for a scalar theory in the standard momentum
representation to leading order in the derivative expansion.  The
spectral values are $\lambda(\bar\phi,\bar\phi)=p^2
+U_k^{(2,0)}[\bar\phi,\bar\phi]$, and the measure and the spectral
density in $d$ dimensions are $d\lambda\, \rho(\bar\phi;\lambda) =
\s012 d p^2 (p^2)^{d/2-1}/(2\pi)^{d/2}$ with density
\begin{eqnarray}\label{eq:0der}
  \rho(\bar\phi,\lambda)=\012 \0{1}{ (2\pi)^{d/2}} 
  (\lambda-U_k^{(2,0)})^{d/2-1}\,
  \theta[\lambda-U_k^{(2,0)}] \,, 
\end{eqnarray}
where $U_k^{(2,0)}=U_k^{(2,0)}[\bar\phi,\bar\phi]$. Convexity is
proven by showing that the spectral values for $\Gamma_k^{(2,0)}+R_k$
are positive for all $k$.  To that end, we first study
\eq{eq:specfullflow} within an additional approximation for the
remaining matrix element. Then we extend the proof to the general
case. For $\bar\phi=\phi$ the spectral representation simplifies
\begin{equation}
  \partial_t\Gamma_k[\phi]=\s012 \int_\R d\lambda\,\rho(\phi;\lambda)\,
  \0{\partial_t R_k
    + \partial_t\lambda(\phi;\lambda)\partial_\lambda R_k}{
    \lambda+R_k(\lambda)}
  \,.
\label{eq:specdiag}
\end{equation} 
In \eq{eq:specdiag} we have defined
$\Gamma_k[\phi]=\Gamma_k[\phi,\phi]$, which only depends on one field.
We have also used that
\begin{eqnarray}\label{eq:flowofla}
  \langle \psi_\lambda |\partial_t 
  \Gamma_k^{(2,0)}
  |\psi_\lambda\rangle=\partial_t \langle \psi_\lambda 
  |\Gamma_k^{(2,0)}|\psi_\lambda\rangle=\partial_t \lambda(\phi;\lambda)\,, 
\end{eqnarray} 
for $\lambda\neq 0$ and $\partial_t \lambda(\phi;\lambda)=\partial_t
\lambda(\phi,\phi;\lambda)$. In \eq{eq:flowofla} we have used that
$\langle
\partial_t \psi_\lambda | \psi_\lambda\rangle=0$ for normalised
functions with $\langle \psi_\lambda |\psi_\lambda\rangle=1$ and
$\Gamma_k^{(2,0)}[\phi,\phi]|\psi_\lambda\rangle=\lambda
|\psi_\lambda\rangle$. The simplicity of the spectral flow
\eq{eq:specdiag} was payed for with the fact that it is not closed
\cite{Litim:2002xm}: the field-dependent input on the rhs, $\rho$ and
$\lambda$ require the knowledge of $\Gamma_k^{(2,0)}[\phi,\phi]\neq
\Gamma_k^{(2)}= \Gamma_k^{(2,0)}+2\Gamma_k^{(1,1)}+ \Gamma_k^{(0,2)}$.
Hence the simplicity of \eq{eq:specdiag} can only be used with the
approximation \cite{Pawlowski:2001df}
\begin{eqnarray}\label{eq:approx1}
  \Gamma_k^{(2)}[\phi]=\Gamma_k^{(2,0)}[\phi,\phi]\,.   
\end{eqnarray} 
Within this truncation \eq{eq:specfullflow} turns into a closed flow
equation for $\Gamma_k[\phi]$. The spectral values are given by
$\lambda(\phi)=\langle\psi_\lambda |\Gamma^{(2)}[\phi] |
\psi_\lambda\rangle$. The flow \eq{eq:specdiag} with \eq{eq:approx1}
allows for the construction of gauge invariant flows
\cite{Pawlowski:2001df,back,SBLiaogauge}. \smallstep

If also neglecting the contributions in \eq{eq:specdiag} that are
proportional to $\partial_t\lambda$, we are led to the widely used
proper-time flows, see \cite{Litim:2002xm}, with spectral
representation
\begin{eqnarray}\label{eq:specflow}
  \partial_t\Gamma_k[\phi]=\s012 \int_\R d\lambda\,\rho(\phi;\lambda)\,
  \0{\partial_t R_k(\lambda)}{\lambda+R_k(\lambda)} \,. 
\end{eqnarray}
The only $\phi$-dependence in \eq{eq:specflow} is that of
$\rho(\phi;\lambda)$ as $\lambda$ serves as an integration variable.
In distinction to the full flow \eq{eq:specfullflow}, we stress that
convexity for the proper-time flow \eq{eq:specflow} is not
automatically guaranteed by formal properties of the effective action.
The flow \eq{eq:specflow} relies on the approximation \eq{eq:approx1},
and $\Gamma_k[\phi]$ is not directly defined as a Legendre transform.
Hence proving convexity for proper-time flows further sustains its
nature as a well-controlled approximation of functional flows.
Indeed, the representation \eq{eq:specflow} facilitates the analysis.
Proving convexity from the flow itself is more difficult for the full
flow, even though we know on general grounds that it entails
convexity. \step

\noindent {\it Convexity of proper-time flows.---} 
If $\Gamma_k^{(2)}$ has negative spectral values they are bounded from
below. Hence, the spectral density obeys
$\rho(\phi;\lambda<\lambda_{\rm min})\equiv 0$ for some finite
$\lambda_{\rm min}$ for all $\phi$. The flow
$\partial_t\Gamma^{(2)}_k[\phi]$ entails the flow of the spectral
values $\lambda(\phi)$ and, in particular, that of $\lambda_{\rm
  min}$. We shall prove that with $k\to 0$ the flow increases
$\lambda_{\rm min}$, its final value being $\lambda_{\rm min}(k=0)\geq
0$. The flow of $\lambda$ is derived from \eq{eq:specflow} with
\eq{eq:flowofla} and \eq{eq:approx1}. The field derivatives only hit
$\rho$ on the rhs of \eq{eq:specflow} and we arrive at
\begin{eqnarray}\label{eq:exflowofla}
  \partial_t\lambda(\phi)=\s012 \int_\R d\lambda'\, 
  \langle \rho^{(2)}(\phi;\lambda')\rangle_{\lambda}\,
  \0{\partial_t R_k(\lambda')}{
    \lambda'+R_k(\lambda')}\,, 
\end{eqnarray}
with $\langle \rho^{(2)}\rangle^{\ }_{\lambda} =
\langle\psi_\lambda|\rho^{(2)}|\psi_\lambda\rangle$.  For the standard
class of regulators used for proper-time flows
\cite{Litim:2002xm}, 
the flow reads
\begin{eqnarray}\label{eq:flowmofla}
  \partial_t\lambda(\phi)= \int_\R d\lambda'\, 
  \langle \rho^{(2)}(\phi;\lambda')\rangle_{\lambda}\,
  \0{1}{
    \left(1+\lambda'/(m k^2)\right)^m}\,.
\end{eqnarray}
Using \eq{eq:0der}, a simple example for $\rho^{(2)}$ is
provided by the leading order derivative expansion in $d=4$,
\begin{eqnarray}\nonumber 
  &&\hspace{-1.2cm} 
  \langle \rho^{(2)}\rangle_{\lambda}=-\0{1}{ (8\pi^2)} \Bigl(U_k^{(4)}\, 
  -2(U_k^{(3)})^2\partial_\lambda\\ 
  &&\hspace{.2cm}  -(\lambda-U_k^{(2)})\,
  (U_k^{(3)})^2 \partial^2_\lambda\Bigr)\theta[\lambda-U_k^{(2)}]\,. 
  \label{eq:coupling} \end{eqnarray}
We proceed by evaluating \eq{eq:exflowofla} for $\lambda_{\rm min}$.  To that
end we have to choose $\phi_0$ that admit the spectral value $\lambda_{\rm
  min}$. Note that the spectral density (and its derivatives) may vanish in
more than two dimensions, $\rho(\phi_0;\lambda_{\rm min})= 0$, $e.g.$~in the
above example of the derivative expansion with $\lambda_{\rm min}=
U^{(2,0)}[\phi_0,\phi_0]$, see \eq{eq:0der}. Moreover, the proofs below work
if no discrete set of low lying spectral values is present, such as come about
in theories with non-trivial topology. However, it can be easily extended to
this case as these modes can be separated due to their discreteness. Assume
that $\lambda_{\rm min}$ stays negative in the limit $k\to 0$. Then, the
propagator generically develops a singularity at the minimal spectral value
at some cut-off scale $k_{\rm sing}$,
\begin{eqnarray}\label{eq:ksing}
  R_{k_{\rm sing}}(\lambda_{\rm min})=-\lambda_{\rm min}\,. 
\end{eqnarray} 
For example, \eq{eq:ksing} holds for (smooth) regulators with
$R_{k=0}\equiv 0$. In \eq{eq:ksing} we have deduced from the
parameterisation $R_k(\lambda)=\lambda\,r(\lambda/k^2)$ and continuity
that the singularity is developed at $\lambda_{\rm sing}$. Later we
shall also discuss the general case.  The contribution of the vicinity
of the singularity dominates the integral if the singularity is strong
enough. We use that $\rho(\phi;\lambda_{\rm min})$ and
$\rho^{(2)}(\phi;\lambda_{\rm min})$ vanish for $\phi$ that do not
admit the eigenvalue $\lambda_{\rm min}$. Consequently as operator
equations we have
\begin{eqnarray}\label{eq:mineqs}
  \rho^{(1)}(\phi;\lambda_{\rm min})\equiv 0,\qquad 
  \rho^{(2)}(\phi;\lambda_{\rm min})\leq 0\,, 
\end{eqnarray} 
in particular for $\phi=\phi_0$. The second identity follows within an
expansion about $\phi_0$ since the related term has to decrease the
spectral density. With \eq{eq:mineqs} the rhs of \eq{eq:exflowofla} is
negative
\begin{eqnarray}\label{eq:signflow}
  \partial_t \lambda_{\rm min}  
  \leq 0\, ,
\end{eqnarray}
and $\lambda_{\rm min}$ is increased for decreasing $k$. As long as
$\Gamma_k$ is differentiable w.r.t.\ $\phi$ this argument applies also
for eigenvalues in the vicinity of $\lambda_{\rm min}$. \smallstep

The condition \eq{eq:signflow} is necessary but not sufficient for
convexity. A sufficient condition is given by the positivity of the
gap $\epsilon=\lambda_{\rm min}+R_k(\lambda_{\rm min})$. Hence, for
$\epsilon\to 0$ its flow $\partial_t \epsilon$ has to be negative.
This leads to the constraint
\begin{eqnarray}\label{eq:epsflowprop}
  \partial_t \lambda_{\rm min}\leq -\left.\0{\partial_t R_k}{1+
    \partial_\lambda R_k}\right|_{\lambda_{\rm min}}\,,
\end{eqnarray} 
as $\partial_t R_k\geq 0$ implies $1+\partial_\lambda R_k\geq 0$. At
$k_{\rm sing}$ an upper bound for $\partial_t \lambda_{\rm min}$ is
obtained from \eq{eq:exflowofla} with $\rho^{(2)}\propto
(\lambda-\lambda_{\rm min})^{\alpha_\rho}$, where we count $\delta(x)$
as $x^{-1}$.  The exponent is bounded from above,
$\alpha_\rho\leq d/2-2$. This follows from the positivity of the
anomalous dimension of the two point function, $\alpha>0$ with
$\lambda-\lambda_{\rm min}\propto p^{2(1+\alpha)}$, and $\rho\propto
p^{2(d/2-1)}$. Negative $\alpha$ would entail a diverging
$\partial_{p^2}\lambda_{\rm min}$ which can only be produced from a
diverging flow $\partial_t\partial_{p^2}\lambda_{\rm min}|_{k_{\rm
    sing}}$.  However, for $\alpha<0$ this flow is finite due to the
suppression factor $\rho^{(2)}$ and $\alpha>0$ follows, for all $k$.
We expand the integrand in \eq{eq:exflowofla} about $\lambda_{\rm
  min}$ as
\begin{eqnarray}
  \0{\partial_t R_k(\lambda)}{\lambda+R_k(\lambda)} 
  =\0{c_1}{\epsilon^\delta+ c_2\,  
    (\lambda-\lambda_{\rm min})^\beta} +{\rm sub\mbox{-}leading}\,.
\label{eq:expansion}\end{eqnarray}
with expansion coefficients $c_1, c_2$. The sub-leading terms comprise
higher order terms in $\epsilon$ and in $(\lambda-\lambda_{\rm min})$.
The exponents $\delta(R_k),\beta(R_k)$ are regulator-dependent real
positive numbers, and essential singularities are covered by the limit
$\delta,\beta\to \infty$, $e.g.$~\eq{eq:flowmofla} with
$\beta=m\to\infty$.  In the latter case the essential singularity is
obtained at $k_{\rm sing}=0$, and the regulator
$R_{k=0}(\lambda)=-\lambda$ for $\lambda<0$.  We conclude that a
sufficient growth of $\lambda_{\rm min}$ is guaranteed for
$\beta\geq d/2-1$ which is identical with $m\geq d/2-1$ in
\eq{eq:flowmofla}.  Lower $m$ correspond to flows for $\Gamma^{(2)}$
with UV problems, in particular the Callan-Symanzik flow for $m=1$ in
$d\geq 4$, whereas the above constraint comes from an IR
consideration: for the flows \eq{eq:flowmofla} UV
finiteness of the flow and the demand of an IR singularity for the
flow of $\lambda_{\rm min}$ are the same, as the flows are monomials
in the propagator. For general regulators there is no UV-IR
interrelation. For $\beta\geq d/2-1$ it follows from
\eq{eq:exflowofla} that 
\begin{eqnarray}\label{eq:estimate}
  \lim_{\epsilon\to0}\partial_t\lambda_{\rm min}= -\infty\,, 
\end{eqnarray}
satisfying \eq{eq:epsflowprop} for $\partial_\lambda R(\lambda_{\rm
  min})>-1$.  In \eq{eq:estimate} we have used that for small enough
$\epsilon$ the integral is dominated by the vicinity of the pole where
$\langle\rho^{(2)}(\phi;\lambda)\rangle_{\lambda_{\rm min}}\leq 0$.
For small enough $\epsilon$ the flow \eq{eq:estimate} exceeds the
decrease of $R_k$, and the singularity cannot be reached.  We conclude
that $\lambda_{\rm min}+R_k(\lambda_{\rm min})>0$ and consequently
\begin{eqnarray}\label{eq:convexity}
  \lim_{k\to 0}\lambda_{\rm min} \geq 0\,, 
\end{eqnarray}
which entails convexity for proper-time flows.  Let us also study the
convexity of truncations to \eq{eq:specflow}: the arguments above
straightaway applies to truncations $\Gamma_{\rm trunc}$ which admit
the direct use of the full field-dependent propagator $(1+\Gamma_{\rm
  trunc}^{(2)}[\phi]/(m\, k^2))^{-1}$ in \eq{eq:specflow}.  If
expansions $\Gamma_{\rm trunc}=\Gamma_1+\Delta\Gamma$ are used on
the rhs of \eq{eq:specflow} (leading to
$(1+\Gamma_{1}^{(2)}[\phi]/(m\, k^2))^{-1}$), convexity might become a
difficult problem.  Then, the arguments above entail convexity of
$\Gamma_1$ for $k\to 0$ but not necessarily for $\Gamma_{\rm trunc}$.
We close with the remark that for $\beta< d/2-1$ convexity cannot be
proven.  Indeed it can be shown that then convexity is not guaranteed
for $k=0$ \cite{DJL}. This holds true for full flows within lowest
order derivative expansion \cite{Litim}. In the latter case it hints 
at inappropriate initial conditions. \smallstep

For regulators that do not lead to singularities \eq{eq:ksing} in the
propagator necessarily $R_{k=0}(\lambda)> |\lambda|$ for $\lambda<0$,
and convexity of $\Gamma_{k=0}$ cannot be guaranteed. Note also, that
for non-convex effective action we keep an explicit regulator dependence for
$k=0$. \step

\noindent {\it Convexity of full flows and general theories.---} The
flow on $\partial_t\lambda$ is given by the second derivative w.r.t.\
$\phi$ of \eq{eq:specfullflow} at $\bar\phi=\phi$. We evaluate the
flow at $\phi=\phi_0$ with minimal spectral value $\lambda_{\rm
  min}(\phi_0)$, and in the vicinity of the singularity, $\lambda_{\rm
  min}+R_k(\lambda_{\rm min})=\epsilon$. We are led to
\begin{eqnarray}\nonumber 
  &&\hspace{-.7cm}\partial_t \lambda_{\rm min} =\s012 
  \int_\R d\lambda'\,\rho(\phi_0;\lambda') \\\nonumber 
  && \hspace{-.3cm} 
  \times 
  \langle\, \langle\psi_{\lambda'}|
  \left(\0{1}{\Gamma^{(2,0)}_k+R_k}\right)^{(2,0)}|\psi_{\lambda'}\rangle
  \,\rangle^{\ }_{\lambda_{\rm min}}\\
  &&\hspace{-.3cm} \times \Bigl[ \partial_t R_k(\lambda') 
  +\partial_t\lambda(\phi_0;\lambda') 
  \partial_{\lambda'}R_k(\lambda')\,\Bigr]+\Delta\,, 
  \label{eq:fullflowla} \end{eqnarray} 
where $\Delta$ comprises sub-leading terms that are proportional to 
off-diagonal matrix elements of the propagator. For $\beta\geq d/2-1$ these 
terms are suppressed by higher order in $\epsilon$. 
There are no terms proportional to $\rho^{(2)}$ and 
$\partial_t \langle\lambda(\phi)^{(2)} 
\rangle_{\lambda_{\rm min}}$ as $\rho$, $\partial_t\lambda$ and 
$|\psi_\lambda\rangle$ only depend on $\bar\phi$. 
All terms in \eq{eq:fullflowla} are proportional to the diagonal 
matrix elements in the second line. Similarly as for
$\rho^{(2)}$ it also follows that the relevant diagonal matrix element
in the integral in \eq{eq:fullflowla} is negative in the vicinity of
$\lambda_{\rm min}$: the propagator takes its maximal spectral value at 
$\phi_0$ and hence its second field
derivative at $\phi_0$ is negative. We conclude for $\beta\geq d/2-1$
that \eq{eq:fullflowla} is only solved for $\partial_t
R_k+\partial_t\lambda\,\partial_\lambda R_k\to 0$ for $\lambda\to
\lambda_{\rm min}$. This entails that $\partial_t \epsilon
=\partial_t\lambda_{\rm min}+{\rm sub\mbox{-}leading}$, and leads to
\begin{eqnarray}\label{eq:epsflow} 
\partial_t \epsilon=-\left.\0{\partial_t R_k}{\partial_\lambda R_k}
\right|_{\lambda_{\rm min}}+{\rm sub\mbox{-}leading}\,.
\end{eqnarray}
The flow of the gap $\epsilon$ has to be negative for $\epsilon\to 0$ 
in order to ensure convexity. This leads to the constraint 
\begin{eqnarray}\label{eq:constraint} 
\left.\0{\partial_t R_k}{\partial_\lambda R_k}
\right|_{\lambda_{\rm min}}\geq 0\,, 
\end{eqnarray}
for small enough $\epsilon$. We remark that \eq{eq:constraint} cannot
hold for all $\lambda$ as $R_k$ has to decay for large positive
$\lambda$, and has to vanish for $k\to 0$. Furthermore the above proof
at $\phi=\bar\phi$ is sufficient for convexity for all $\phi$. If
evaluating the full flow at some $\bar\phi\neq \phi_0$ the spectral
density is non-vanishing at this $\bar\phi$ and we get convexity for
$\beta\geq 1$. This completes the convexity proof of general flows.

The proof is straightforwardly extended to theories with general field
content with fields $\phi_i$, $i=1,...,N$. For illustration we
restrict ourselves to regulators that are diagonal in field space with
entries $R^{(1)}_k,...,R^{(N)}_k$ and arguments $\Gamma^{(2,0)}_{k,ii}$,
the diagonal elements of the two-point function. Choosing a spectral
representation in terms of the eigenfunctions $\psi_\lambda^{(i)}$
and spectral values $\lambda^{(i)}$ of $\Gamma^{(2,0)}_{k,ii}$, the
integrand in \eq{eq:specfullflow} reads
\begin{eqnarray}\label{eq:integrand} 
\sum_{i=1}^N\rho_i(\bar\phi;\lambda)
  \langle \psi^{(i)}_\lambda|\left(\0{1}{
    \Gamma_k^{(2,0)}+R_k}\right)_{ii}\partial_t R^{(i)}_k|
  \psi^{(i)}_\lambda\rangle\,. 
\end{eqnarray} 
Note that the spectral values $\lambda^{(i)}$ are in general not
spectral values of $\Gamma_k^{(2)}$. However, singularities of
diagonal elements of the propagator in \eq{eq:integrand} are in one to
one correspondence to vanishing spectral values of diagonal elements
of the two point function, $\Gamma_{k,ii}^{(2,0)}+R_k^{(i)}$. Hence,
$\lambda^{(i)}\geq 0$ at $k=0$ for all $i$ follows directly from the
proof for theories with only one field, and it entails $\lambda\geq
0$, where $\lambda$ are the spectral values of $\Gamma^{(2,0)}$ at
$k=0$. \step

\noindent {\it Derivative expansion.---} 
To illustrate our findings, we consider the infrared running of the
scale-dependent effective potential $U_k(\phi)$ in $d=3$ dimensions for a
$N$-component real scalar field $\phi^a$ in the large-$N$ limit, to leading
order in a derivative expansion, $e.g.$~\cite{Tetradis:1992qt}.
\begin{figure}[h]
\vspace{.3cm}
\begin{picture}(50,210)(0,0)
 \put(-85,90){\epsfig{file=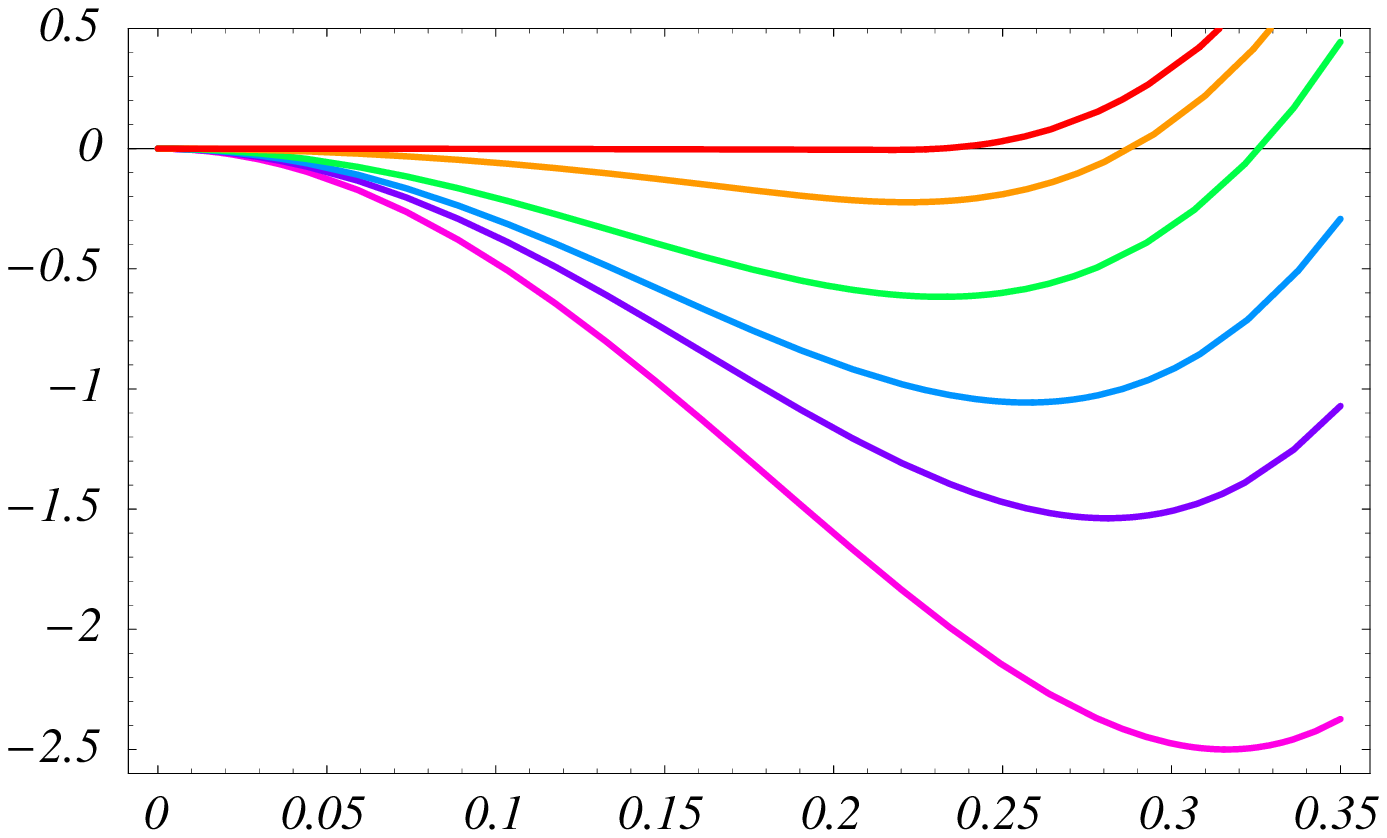,width=
.9\hsize
}}
\put(-22,145){\large $U_k(\phi)$}
\put(-85,192){\large $a)$}
\put(128,95){\large $ \phi$}
\put(-92,15){\epsfig{file=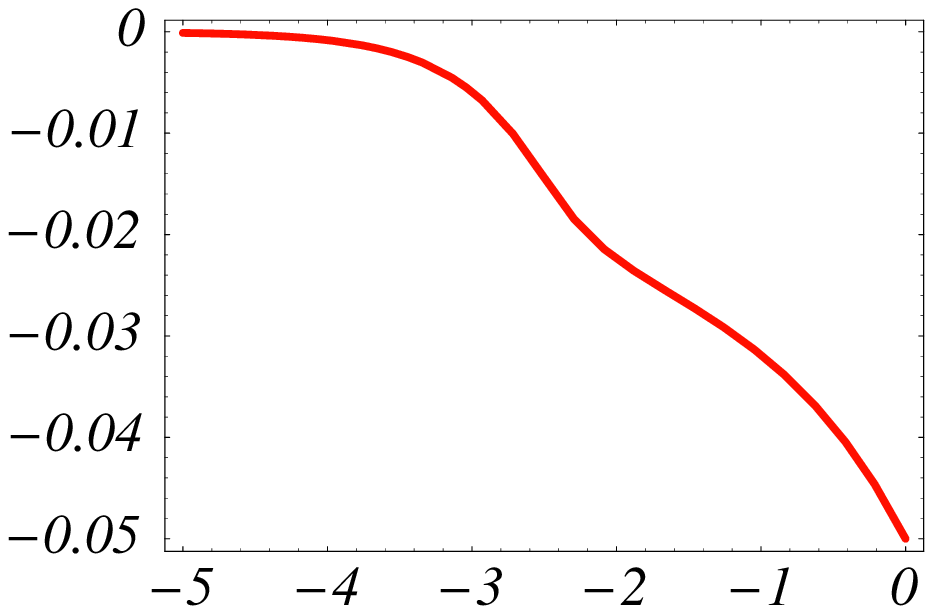,width=.47\hsize}}
\put(27,15){\epsfig{file=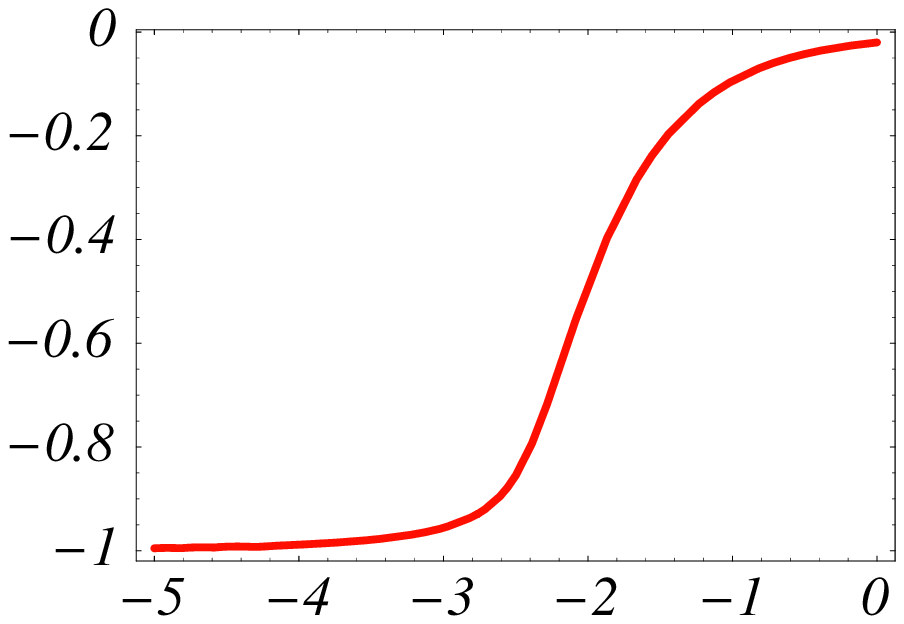,width=.47\hsize}}
\put(-85,82){\large $b)$}
\put(-48,45){ $\displaystyle\frac{\lambda_{{\rm min}}}{\Lambda^2}$}
\put(22,13){$t$}
\put(30,82){\large $c)$}
\put(70,45){ $\displaystyle\frac{\lambda_{{\rm min}}}{k^2_{\rm eff}}$}
\put(140,13){$t$}

\end{picture}
\vspace{-.8cm}
\caption{Approach to convexity in terms of $a)$ the effective
  potential (rescaled) for $t=\ln k/\Lambda=0,-0.5,-1,-2,-3,-5$ from
  bottom to top, and $b,c)$ the lowest spectral value $\lambda_{{\rm
      min}}$; $d=3$, $\mu^2/\Lambda^2=-0.05$,
  $g/\Lambda=1$ (see text).}\label{fig:05pot} \vspace{-.5cm}
\end{figure} Here, the running potential $U_k$ is obtained from integrating
the proper-time flow \eq{eq:specflow} in the parametrisation \eq{eq:flowmofla}
with $m=d/2+1$, see Fig.~\ref{fig:05pot}$a)-c)$.  This value of $m$
corresponds to an optimised flow \cite{Litim:2001up,Pawlowski:2005xe}, similar
plots follow for all $m\geq 3/2$ \cite{DJL}. The boundary condition is
$U_\Lambda=\frac{1}{2}\mu^2\phi^2+\frac{1}{8}g\phi^4$ at $k=\Lambda$.  For
$\mu^2/g<0$, the potential $U_\Lambda$ displays spontaneous symmetry breaking
with a global minimum at $\phi^2_{{\rm min},\Lambda}=-2\mu^2/g$.  With
decreasing $k$, the minimum runs towards smaller values, settling at
$\phi_{{\rm min},0}<\phi_{{\rm min},\Lambda}$, see Fig.~\ref{fig:05pot}$a)$.
For fields in the non-convex regime of the potential the flow displays
negative spectral values, corresponding to an instability. Here, the lowest
spectral value is given by the running mass term at vanishing field,
$\lambda_{{\rm min}}=U_k''(0)\le 0$, which smoothly tends to zero for $k\to
0$, see Fig.~\ref{fig:05pot}$b)$. Once $\phi_{{\rm min}}$ has settled, 
the running of $\lambda_{{\rm min}}$ changes
qualitatively: in the infrared, the size of the spectral value is set by the
effective cutoff scale $k^2_{\rm eff}(k)=mk^2$, see Fig.~\ref{fig:05pot}$c)$,
and the entire inner part of the potential becomes convex.\step

\noindent {\it Discussion.---} We have provided a proof of convexity
for general functional flows \eq{eq:specfullflow}, subject to simple
constraints on the set of regulators. The constraints are $\beta\geq
d/2-1$ derived from \eq{eq:expansion}, as well as \eq{eq:constraint}
for full flows.  The finiteness of $\partial_t \lambda_{\rm min}$ at
the singularity and \eq{eq:constraint} seemingly indicates worse
convexity properties for the full flow \eq{eq:specfullflow} (at
$\phi=\phi_0)$ in comparison with proper-time flows. However, full
flows entail convexity by definition. This paradox is resolved by
considering the initial condition. Only consistent choices correspond
to a path integral and lead to convex effective actions at $k=0$.
Hence, regulators that violate \eq{eq:constraint} can be used to test
the consistency of initial conditions for $\Gamma_k$ for full flows.
This allows us to investigate physical instabilities within these
settings.  \smallstep

In addition we have proven positivity of the infrared anomalous dimension of
the propagator, $\alpha\geq 0$. Negative $\alpha$ require additional fields
with at least one strictly positive anomalous dimension. The latter scenario
is relevant $e.g.$~for Landau gauge QCD, where \cite{Pawlowski:2003hq} already
anticipates the general result.  \smallstep

The present work also finalises the analysis initiated in
\cite{Litim:2001hk,Litim:2002xm}, and fully establishes
proper-time flows as well-defined, convexity-preserving approximations
of first-principle flows.  Note that in the proper-time approximation
the standard regulators leading to \eq{eq:flowmofla} violate
\eq{eq:constraint}. For stable flows beyond \eq{eq:specflow} one
should modify these regulators for negative spectral values.
\smallstep

{\it Acknowledgements:}\ 
We acknowledge Fondecyt-Chile grants No.1020061 and
7020061, and DFG support under contract GI328/1-2.  The work of DFL is
supported by an EPSRC Advanced Fellowship.

\end{document}